\documentclass[3p,final,11pt,authoryear]{elsarticle}
\usepackage[colorlinks]{hyperref}
\usepackage[utf8]{inputenc}
\usepackage{graphicx}
\usepackage{mathtools}
\usepackage{amsmath}
\usepackage{lscape}
\usepackage[nameinlink,noabbrev]{cleveref}
\usepackage{enumitem}
\usepackage{natbib}
\usepackage{subcaption}
\usepackage[tableposition = top]{caption}
\usepackage{booktabs}
\usepackage{algorithm}
\usepackage{algorithmic}
\usepackage{hyperref}
\usepackage{float}
\usepackage{tabularx}
\usepackage{adjustbox}
\usepackage{rotating}
\usepackage{lscape}
\usepackage{amsfonts}
\usepackage{placeins}

\setcitestyle{authoryear,open={(},close={)}}
\usepackage{setspace}

\begin{document}

\begin{frontmatter}

\title{Traffic Collisions: Temporal Patterns and Severity-Weighted Hotspot Analysis} 

    \author[1]{Nael Alsaleh\corref{cor1}}\ead{nael.alsaleh@aurak.ac.ae}
    \author[1]{Noura 
    Falis}\ead{2023006419@aurak.ac.ae}
    \author[2]{Tareq Alsaleh}\ead{talsaleh@torontomu.ca}
    \author[1]{Farah 
    Ba Fakih}\ead{farah.bafakih@aurak.ac.ae}

    \address[1]{Department of Architecture and Civil Engineering, American University of Ras Al Khaimah, Ras Al Khaimah 72603, United Arab Emirates}
    \address[2]{Laboratory of Innovations in Transportation (LiTrans), Toronto Metropolitan University, Canada}
    \cortext[cor1]{Corresponding Author.}

\begin{abstract}

Understanding traffic collision patterns is of high importance for effective road safety planning in fast-growing urban environments. This study examines the temporal and spatial patterns of traffic collisions in Dubai, UAE, with a particular focus on collision severity. To this end, traffic collision records from November 2024 to June 2025 were analyzed to examine hourly, daily, and monthly variations in collision frequency and severity for both overall traffic collisions and pedestrian-related accidents. Temporal associations with severity were evaluated using chi-square tests and Cramér’s V, while spatial patterns were analyzed using severity-weighted hotspot analysis based on the Getis–Ord Gi* statistic, complemented by inverse distance weighting (IDW) interpolation. The results show a clear temporal variation in overall collision frequency and severity, with higher collision frequencies during evening and nighttime periods with 44\% higher probability of high-severity outcomes at night compared to the afternoon. On the other hand, pedestrian-related accidents showed a distinct temporal profile, characterized by higher occurrence during late-evening hours and relatively limited variation across days of the week and months. Spatial analysis identified statistically significant severity hotspots for overall collisions in the northern and northwestern parts of Dubai and along the Al Ain–Dubai Highway, while pedestrian severity hotspots were concentrated near industrial areas in the southwestern region. Several policy measures are proposed based on the findings including, reducing nighttime speed limits, enhancing automated enforcement, improving roadway lighting, and implementing pedestrian-focused treatments in statistically significant hotspots.

\end{abstract}

\begin{keyword}
Accident hotspot identification, road traffic accidents, accident severity, Getis–Ord Gi* statistic, temporal and spatial analysis.

\end{keyword}
\end{frontmatter}

\section{Introduction}
\label{sec:Introduction}

Road traffic accidents remain a significant public safety issue, particularly in fast-growing urban environments. Globally, the scale of the problem is substantial, with a reported 1.19 million road traffic deaths in 2021, approximately 21\% of which involved pedestrians \citep{who2023roadsafety}. In the United Arab Emirates (UAE), the total number of traffic accidents in 2023 was estimated at 4,391, with nearly half (49.56\%) occurring in the Emirate of Dubai \citep{uaestat2025traffic}. Although the country has reported improvements in road safety over the years, recent figures indicate that fatalities from traffic accidents have shown a slight upward trend, increasing by approximately 11\% from 343 deaths in 2022 to 384 deaths in 2024. This rise can be largely attributed to sustained population growth, increasing vehicle ownership and higher mobility demands \citep{alamir2025uaeaccidents}.

Understanding the temporal and spatial patterns of traffic accidents is critical for evidence-based planning and targeted interventions. These insights support global road safety targets, including Sustainable Development Goal (SDG) 3.6, which aims to reduce 50\% of traffic deaths and injuries by 2030. The World Health Organization's Decade of Action for Road Safety 2021–2030 calls for continuous improvements in road and vehicle design, safer infrastructure, and stronger post-crash response systems \citep{who2021decadeaction}. Temporal and spatial analyses help identify high-risk periods, hotspots, and vulnerable road-user groups, which enables targeted interventions and optimized resource allocation.

Prior research highlights the wide range of factors that influence traffic accidents across regions. A recent study performed a spatial and temporal analysis of traffic accidents in four major Californian cities, including Los Angeles, Sacramento, San Diego, and San Jose, revealed that the majority of accidents recorded over the five-year analysis period occurred in Los Angeles \citep{alsahfi2024spatial}. Another study demonstrated that in developed urban centers such as Kyoto City in Japan, accident clustering is influenced by urban density, land-use characteristics, and intersection design, with lower crash probabilities observed near parks and higher probabilities near commercial facilities such as restaurants, supermarkets, and convenience stores \citep{nakao2025analysis}. 

In the UAE, careless driving is often the leading factor to traffic accidents. A recent study in Abu Dhabi reported elevated accident intensity, with several hotspots concentrated in the central business district \citep{alkaabi2023identification}. Another study found that drivers under the age of 40 are 18\% more likely than older drivers to be involved in accidents on external roads, including highways \citep{imreizeeq2023statistical}. Additionally, a survey of 458 drivers across the UAE showed that accident occurrence is negatively correlated with driver age and positively correlated with the driving frequency \citep{abuzaid2024driving}.

Previous studies have also demonstrated that inadequate infrastructure and weak enforcement of traffic regulations exacerbate the severity of accidents. For example, studies in Irbid, Jordan \citep{al2025spatial}, and Rawalpindi, Pakistan \citep{aati2024analysis}, reported that vehicle speed has contributed to higher accident severity, which makes effective traffic regulation and enforcement essential for reducing injuries and fatalities. Moreover, narrow lanes and the absence of road markings or traffic signals on Nepal’s Araniko Highway have been associated with a fatality rate of 14.9\%, emphasizing the impact of road design on traffic safety \citep{bhele2024spatial}.

A wide range of analytical approaches have been applied in traffic accident research, pointing to the multidimensional nature of the occurrence of crashes. Machine learning techniques have been extensively used to develop predictive models for accident likelihood and severity due to their ability to capture nonlinear relationships among risk factors (e.g., \citep{al2025spatial, augustine2022road}). Text mining has also emerged as a valuable tool for analyzing narrative crash reports and extracting underlying patterns that are not easily captured through structured data sources (e.g., \citep{jaradat2025investigating}). For spatial hotspot analysis, studies have employed logistic regression, kernel density estimation, and a variety of geospatial statistical techniques. Among these, the Getis-Ord Gi* statistic and Inverse Distance Weighting (IDW) interpolation have been widely adopted and shown to be effective in detecting statistically significant clusters and visualizing the intensity of crash concentrations, offering robust tools for identifying localized areas of elevated risk (e.g., \citep{alkaabi2023identification, hovenden2020use, berhanu2023spatial, hazaymeh2022spatiotemporal}).

Given that traffic accident patterns vary substantially across geographic, socio-economic, and infrastructural contexts, these variations warrant context-specific investigation to support targeted interventions. The present study investigates the temporal patterns and spatial distribution of traffic collision incidents in Dubai, UAE, with emphasis on collision severity. Severity-weighted spatial hotspot analysis is conducted using the Getis-Ord Gi* statistic and IDW interpolation to identify high-risk locations. Temporal analyzes are performed to assess hourly, daily and monthly variations in collision occurrence and severity. All analyses are performed separately for overall collisions and pedestrian-related incidents. This research aims to provide a comprehensive understanding of accident dynamics and to support data-driven road safety planning and targeted interventions.

The remainder of the paper is organized as follows. Section \ref{sec:Data & Methods} explains the analytical framework used to examine traffic collision patterns, including data collection, preprocessing, temporal analysis, and spatial hotspot analysis. Section \ref{sec:Results} presents the temporal and spatial analysis results. Section \ref{sec: recomendations} discusses the study findings and presents key recommendations derived from the results, and Section \ref{sec:conclusions} concludes the paper and provides directions for future work.

\section{Data and Methods}
\label{sec:Data & Methods}
\begin{figure}[!ht]
\centering
\includegraphics[width=0.95\textwidth]{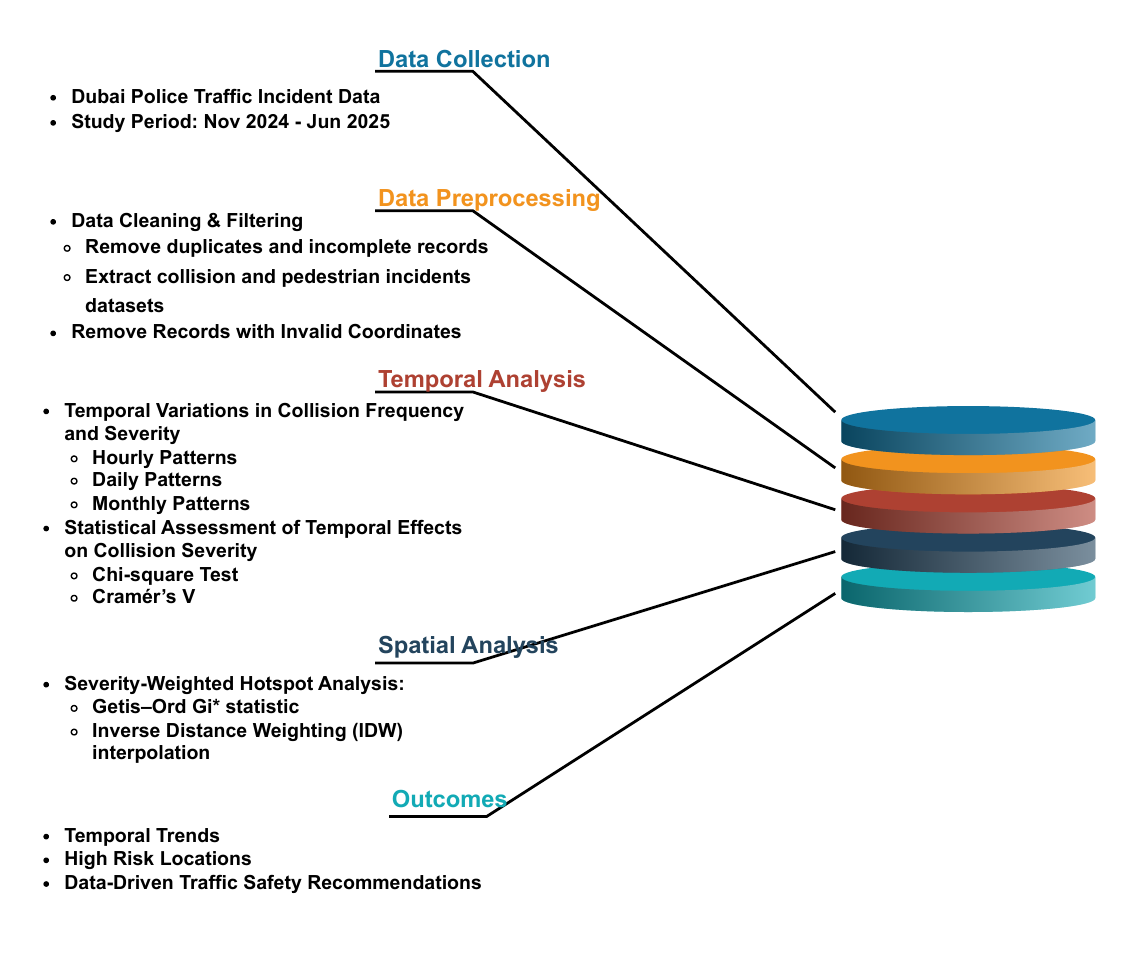}
\caption{Traffic collision analysis framework.}
\label{fig1}
\end{figure}

This section describes the analytical framework used in this study (see Figure~\ref{fig1}) to examine traffic collision patterns in Dubai, including data collection, pre-processing, temporal analysis, and spatial hotspot analysis.

\subsection{Dataset Description}
Traffic incident records were obtained from the Dubai Pulse Open Data Platform \citep{dubaipolice2025traffic}, an official open-data repository maintained by the Government of Dubai. The dataset is published by the Dubai Police Department and comprises reported traffic incidents with associated temporal and spatial attributes. In this study, traffic incident records covering the period from November~5,~2024, to June~2,~2025, were analyzed.

The dataset includes information on incident identification number (ID), date and time of occurrence, descriptive information, and geographic coordinates (longitude and latitude). The description field is provided in Arabic and contains details related to the incident type and severity level. Incident severity is categorized into two levels: Low and High. A total of 71,375 traffic incidents were recorded during the study period. Based on the reported incident type, the records were classified into collision and non-collision incidents.

Collision incidents accounted for approximately 47\% of the total records (33,604 incidents). These incidents include vehicle-object collisions, vehicle-vehicle collisions, pedestrian-related accidents, motorcycle and bicycle collisions, rollover events, animal-related collisions, special vehicle collisions, and hit-and-run incidents. Table~\ref{tab:collision_types} presents a detailed classification of collision subcategories along with their corresponding occurrence counts and percentages.

\begin{table}[ht]
\caption{Classification and frequency of collision incident subcategories}
\centering
\begin{tabularx}{\textwidth}{@{}lXcc@{}}
\hline
\textbf{Collision Subcategory} & \textbf{Brief Description} & \textbf{Count} & \textbf{(\%)} \\
\hline
Vehicle--Object Collisions & Vehicle colliding with a fixed object & 17,149 & 51.03 \\
Vehicle--Vehicle Collisions & Collision between two or more vehicles & 6,901 & 20.54 \\
Motorcycle Collisions & Collision involving a motorcycle & 3,555 & 10.58 \\
Rollover Collisions & Vehicle rollover incident & 2,271 & 6.76 \\
Hit-and-Run & Collision with fleeing party & 1,423 & 4.23 \\
Pedestrian Collisions & Vehicle--pedestrian collision & 1,367 & 4.07 \\
Bicycle Collisions & Collision involving a bicycle & 719 & 2.14 \\
Animal Collisions & Collision involving an animal & 158 & 0.47 \\
Special Vehicle Collisions & Collision with special vehicles (e.g., tram/train or military vehicle) & 61 & 0.18 \\
\hline
\textbf{Total} &  & \textbf{33,604} & \textbf{100.00} \\
\hline
\end{tabularx}
\label{tab:collision_types}
\end{table}

Non-collision incidents represented approximately 53\% of the dataset and included driver behavioral violations (e.g., double parking, street racing, no-entry violations), pedestrian behavioral violations (e.g., unsafe or improper road crossing), infrastructure-related issues (e.g., objects on the roadway, traffic signal malfunctions, oil spills), mechanical issues (e.g., vehicle breakdowns, vehicle fires while in motion, electrical or mechanical faults), and other events such as traffic congestion, flying object impacts, and stray animals on the roadway.

In this study, temporal pattern analysis and spatial hotspot analysis were conducted for all collision incidents and for pedestrian-related accidents using the same analytical framework. Collision incidents and pedestrian-related accidents were selected for analysis due to their direct implications for road user safety and their contribution to traffic-related injuries and fatalities. These analyses aim to identify temporal trends and high-risk locations to support data-driven road safety planning and targeted intervention strategies.

\subsection{Data Pre-processing}
The traffic incident dataset was preprocessed to ensure temporal and spatial consistency prior to analysis. Records were first filtered to retain incidents within the study period (November~5,~2024 to June~2,~2025). The dataset was examined for duplicate entries and missing values. No duplicate records were identified, however, incident data were missing for two days within the study period, specifically November~9 and 10,~2024.

Two analytical datasets were then extracted: all collision incidents and pedestrian-related collision incidents. Spatial validation was performed by visualizing incident locations in ArcMap, verifying their consistency within the administrative boundary of the Emirate of Dubai, and removing records with invalid geographic coordinates. Initially, the collision dataset contained 33,604 records and the pedestrian subset included 1,367 records. After spatial validation, these were reduced to 33,480 and 1,365 records, respectively. The resulting datasets were used for subsequent temporal and spatial analyses.

\subsection{Temporal Analysis}
Temporal analyses were conducted to examine patterns in traffic collisions across different time scales. Collision records were analyzed at hourly, daily, and monthly resolutions to identify variations in collision frequency and severity by time of day, day of the week, and month during the study period. Analyses were performed separately for (i) all collision incidents and (ii) pedestrian related collisions to capture temporal trends associated with different accident types. Associations between temporal characteristics (time-of-day period, day of week, and month) and collision severity were evaluated using the chi-square test of independence, with Cram\'er's $V$ reported as an effect size measure \citep{mchugh2013chi}. The statistical formulation is as follows.

Let each collision record be indexed by $n=1,\dots,N$, with a temporal category $T_n \in \{1,\dots,r\}$ defined by the temporal factor of interest (time of day, day of week, or month), and severity label $S_n \in \{H,L\}$ indicating High ($H$) or Low ($L$) severity. For each temporal factor, an $r \times 2$ contingency table of observed counts $O_{ij}$ was constructed, where $i \in \{1,\dots,r\}$ indexes temporal categories and $j \in \{H,L\}$ indexes severity:
\begin{equation}
O_{ij} \;=\; \sum_{n=1}^{N} \mathbb{I}(T_n=i)\,\mathbb{I}(S_n=j).
\label{eq:obs_counts}
\end{equation}
Under the null hypothesis that $T$ and $S$ are independent, expected counts were computed as
\begin{equation}
E_{ij} \;=\; \frac{\left(\sum_{j} O_{ij}\right)\left(\sum_{i} O_{ij}\right)}{N}.
\label{eq:exp_counts}
\end{equation}
The chi-square statistic was then calculated as
\begin{equation}
\chi^2 \;=\; \sum_{i=1}^{r}\sum_{j \in \{H,L\}} \frac{(O_{ij}-E_{ij})^2}{E_{ij}},
\label{eq:chisq}
\end{equation}
with degrees of freedom
\begin{equation}
\mathrm{df} \;=\; (r-1)(2-1) \;=\; r-1.
\label{eq:df}
\end{equation}
Statistical significance for all tests was evaluated at $\alpha = 0.05$.

Effect size was quantified using Cram\'er's $V$:
\begin{equation}
V \;=\; \sqrt{\frac{\chi^2}{N\,(k-1)}},
\qquad k=\min(r,2)=2,
\label{eq:cramers_v}
\end{equation}
which, for a binary severity outcome, simplifies to
\begin{equation}
V \;=\; \sqrt{\frac{\chi^2}{N}}.
\label{eq:cramers_v_binary}
\end{equation}

For the time-of-day analysis, four periods were defined: Morning (06:00--11:59), Afternoon (12:00--17:59), Evening (18:00--23:59), and Night (00:00--05:59). The same testing procedure was applied for day of week and month using their respective category sets.

\subsection{Spatial Hotspot Analysis}
Spatial hotspot analysis was performed to identify high-risk locations of traffic collisions by accounting for both accident frequency and severity. Each collision was assigned a weight based on its severity level, with low-severity incidents assigned a weight of 1 and high-severity incidents assigned a weight of 2. This weighting scheme can be expressed as
\begin{equation}
w_n =
\begin{cases}
1, & \text{Low severity},\\
2, & \text{High severity},
\end{cases}
\label{eq:severity_weight}
\end{equation}
where $w_n$ denotes the severity weight assigned to collision $n$. The hotspot analysis was conducted separately for all collision incidents and pedestrian-related accidents. The resulting weighted incident dataset was analyzed using the Getis--Ord Gi* statistic and complemented by inverse distance weighting (IDW) interpolation to assess spatial patterns of collision risk.

The Getis--Ord Gi* statistic, originally developed by Getis and Ord \citep{getis1992analysis}, was applied to identify statistically significant spatial clustering patterns at 90\%, 95\%, and 99\% confidence level. This local indicator of spatial association evaluates whether high or low values are spatially concentrated relative to a random spatial distribution. Statistically significant positive Gi* values indicate hotspots, while significant negative values indicate cold spots. Formally, let $x_j$ denote the weighted value at spatial feature $j$ (for example, the severity-weighted count of collisions associated with that feature), and let $w_{ij}$ denote the spatial weight defining the neighbourhood influence between features $i$ and $j$. The standardised Getis--Ord statistic for feature $i$ is given by
\begin{equation}
G_i^* \;=\;
\frac{\sum_{j=1}^{N} w_{ij} x_j \;-\; \bar{x}\sum_{j=1}^{N} w_{ij}}
{S \sqrt{\frac{N\sum_{j=1}^{N} w_{ij}^{2} - \left(\sum_{j=1}^{N} w_{ij}\right)^{2}}{N-1}}},
\label{eq:getis_ord_gistar}
\end{equation}
where $N$ is the number of spatial features, $\bar{x}$ is the mean of $x_j$, and
\begin{equation}
S \;=\; \sqrt{\frac{\sum_{j=1}^{N} x_j^{2}}{N} - \bar{x}^{2}}.
\label{eq:getis_ord_S}
\end{equation}

IDW interpolation was used to generate a continuous surface representing spatial variation in collision intensity. To provide a continuous representation of spatial risk intensity, IDW interpolation was applied to the Gi* outputs. This method estimates values at unsampled locations as a distance-weighted average of nearby observations, with weights inversely proportional to distance. For an unsampled location $s_0$, the IDW estimate can be expressed as
\begin{equation}
\hat{z}(s_0) \;=\; \frac{\sum_{k=1}^{m} z(s_k)\,d(s_0,s_k)^{-p}}{\sum_{k=1}^{m} d(s_0,s_k)^{-p}},
\label{eq:idw}
\end{equation}
where $z(s_k)$ is the input value at sampled location $s_k$ (here, the corresponding $G_i^*$ output), $d(s_0,s_k)$ is the distance between $s_0$ and $s_k$, $m$ is the number of neighbouring points used, and $p>0$ is the distance decay power parameter. Both  Getis--Ord Gi* and IDW interpolation analyses were performed using the Spatial Analyst extension in ArcMap, a desktop GIS platform for spatial data management, analysis, and visualization.

\section{Results}
\label{sec:Results}
This section presents and discusses the temporal analysis of traffic collisions, followed by an analysis of their spatial hotspots.

\subsection{Temporal Analysis}
This subsection examines the temporal characteristics of all collision incidents and pedestrian-related collision accidents.

\subsubsection{All Collision Incidents}

\begin{figure}[!ht]
\centering
\includegraphics[width=0.95\textwidth]{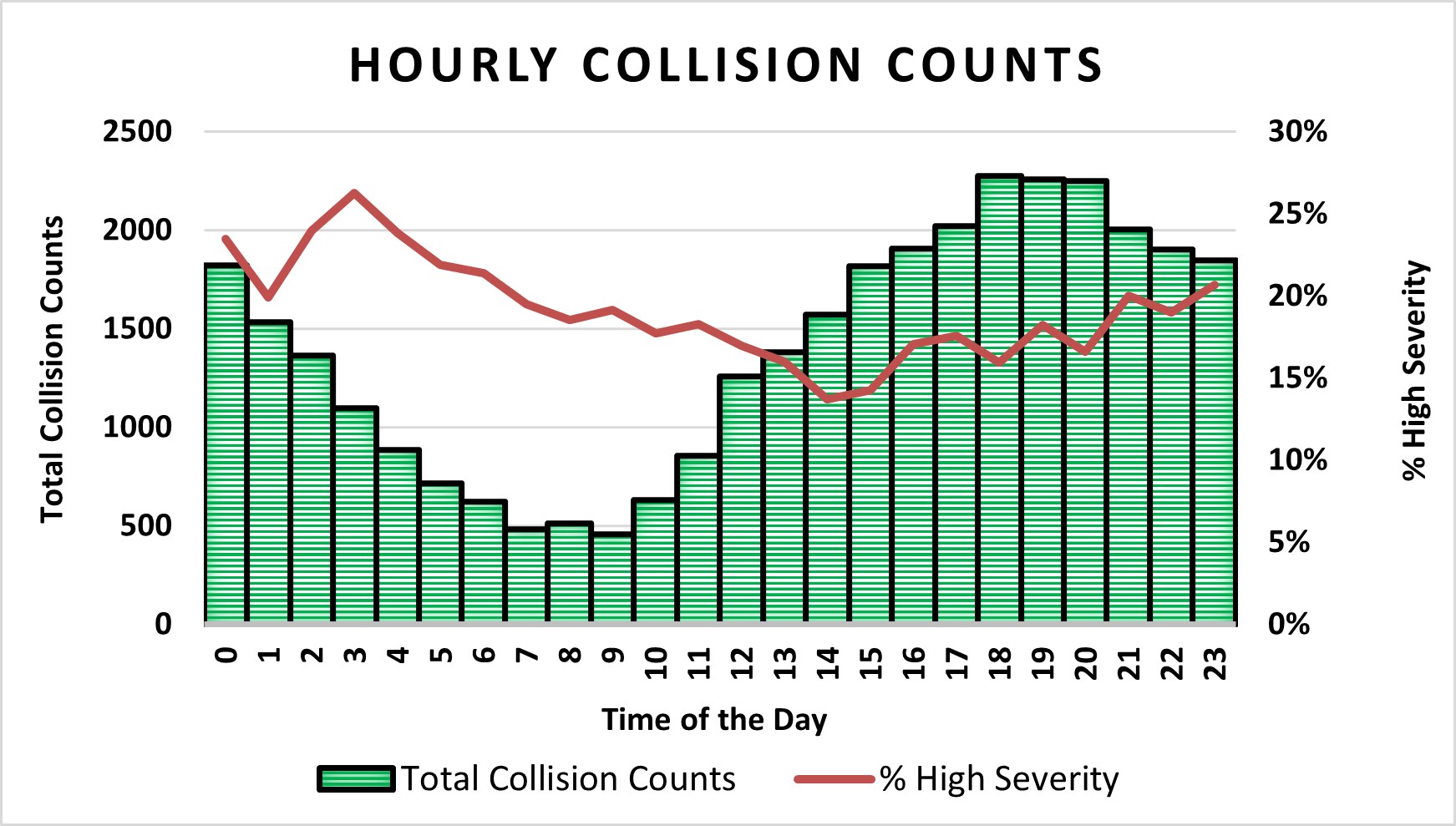}
\caption{Hourly distribution of collision incidents.}
\label{fig2}
\end{figure}

The cleaned collision dataset comprised 33,480 records over the study period, corresponding to an average of 160.96 collisions per day in Dubai. The highest daily collision count was recorded on April 16, 2025, with 252 collisions, whereas the lowest was observed on November 8, 2024, with 52 collisions. The peak hourly collision frequency occurred between 9:00 PM and 10:00 PM on February 18, 2025, during which 26 collisions were recorded. With respect to collision severity, the highest number of high-severity collisions in a single day was observed on November 18, 2024, with 46 severe incidents. To further characterize the temporal distribution of traffic collisions and their associated severity levels, hourly, weekly, and monthly patterns are illustrated in Figures~\ref{fig2} through.~\ref{fig4}.
\FloatBarrier

\begin{figure}[!ht]
\centering
\includegraphics[width=0.95\textwidth]{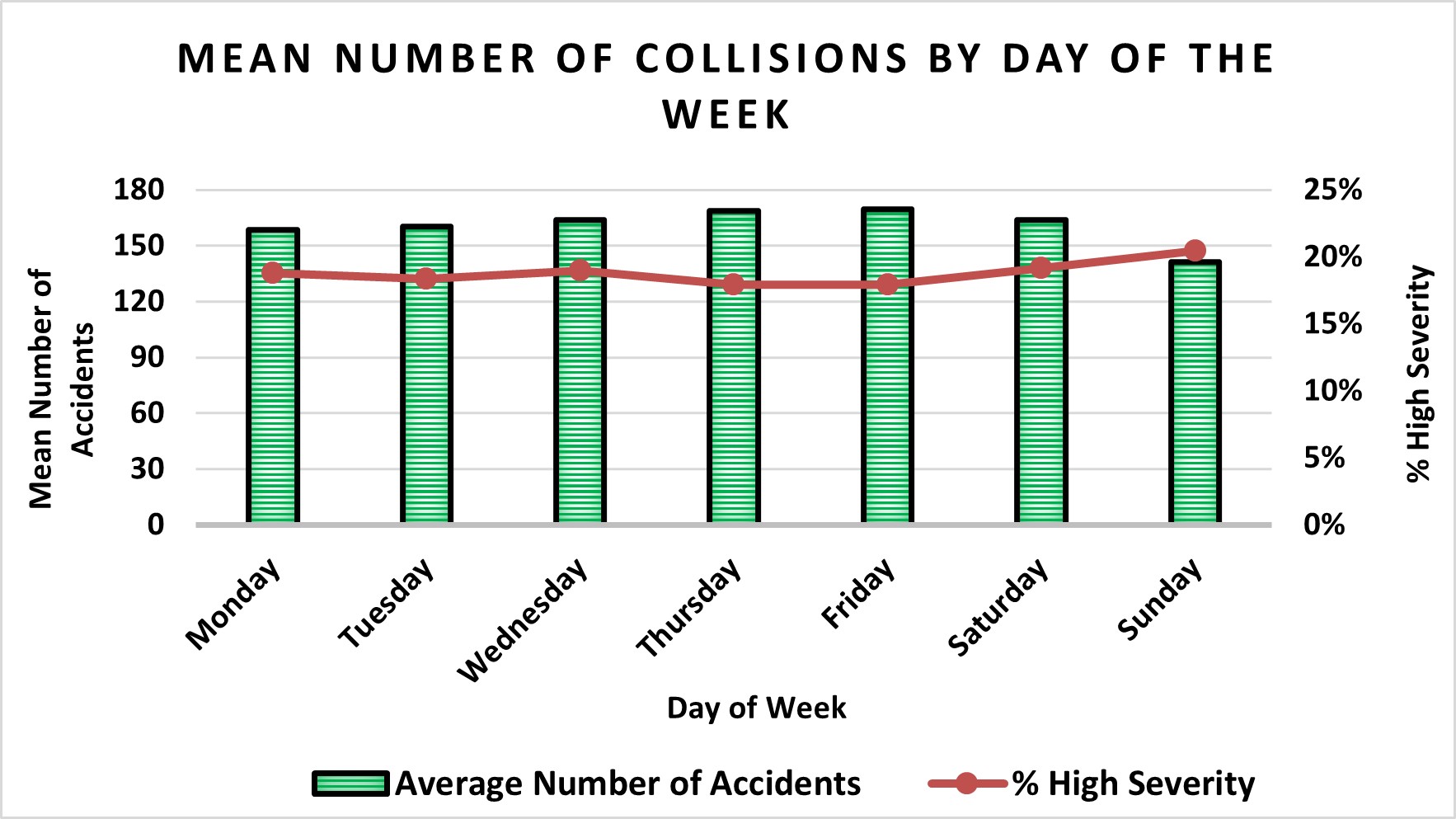}
\caption{Weekly distribution of collision incidents.}
\label{fig3}
\end{figure}

\begin{figure}[!ht]
\centering
\includegraphics[width=0.95\textwidth]{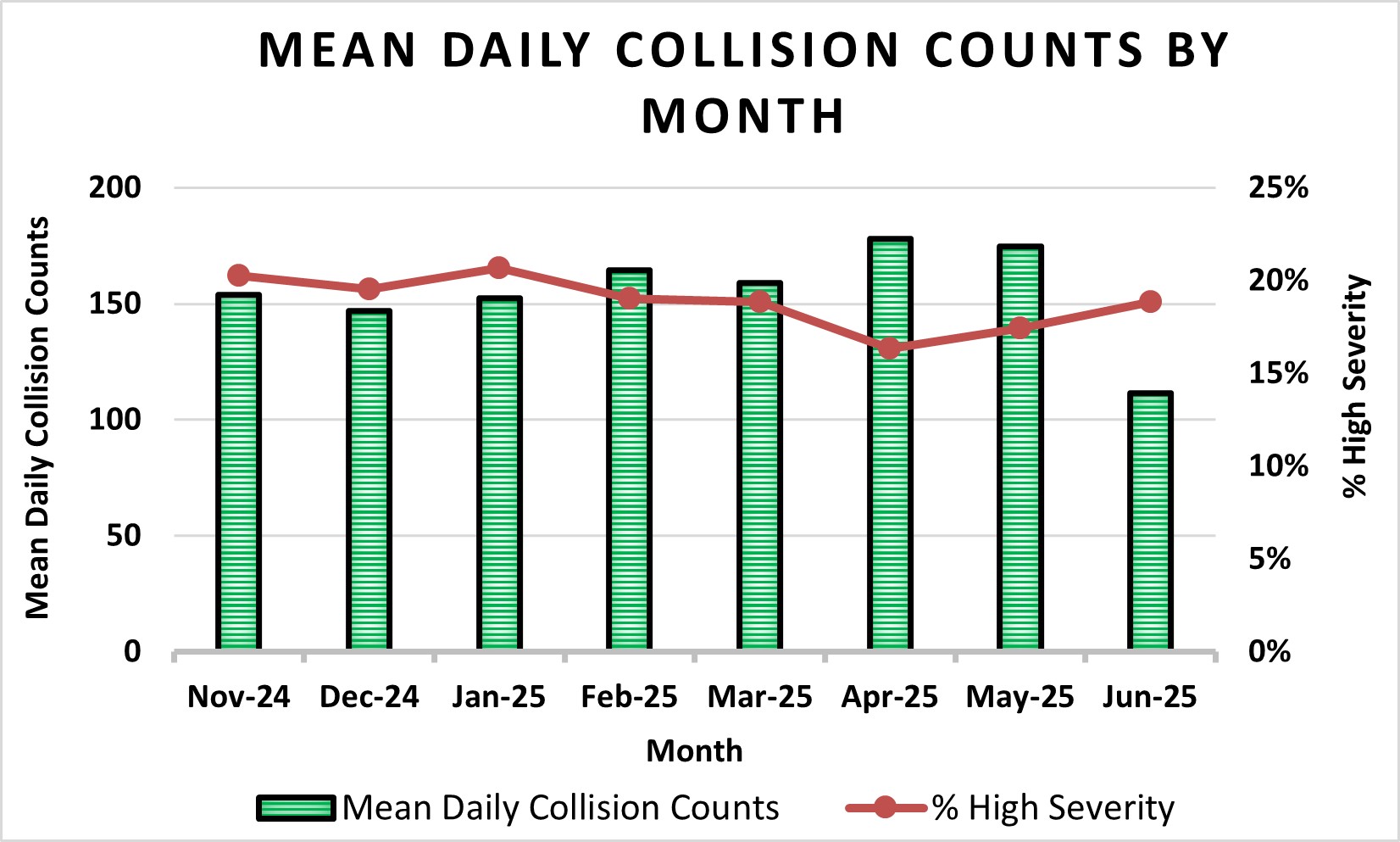}
\caption{Monthly distribution of collision incidents.}
\label{fig4}
\end{figure}

Figure~\ref{fig2} illustrates the hourly distribution of total traffic collisions alongside the percentage of high-severity collisions. Collision counts show a clear diurnal pattern, with lower counts during early morning hours (between 07:00 AM - 09:00 AM) and a steady increase throughout the day, reaching peak levels during the evening period (6:00 PM – 9:00 PM). This pattern corresponds to periods of high traffic demand and activity, and is consistent with findings from previous studies in other urban contexts \citep{al2025spatial, shin2020temporal}.

In contrast, the proportion of high-severity collisions shows a different temporal pattern. Higher percentages of severe collisions are observed during late-night and early-morning hours, despite relatively lower collision counts, while the lowest severity proportions occur during the afternoon. As summarized in Table~\ref{tab:severity_all}, a chi-square test results indicate a statistically significant association between collision severity and time-of-day periods, although the effect size is small (Cramér’s V = 0.066), suggesting that time of day explains a limited but meaningful portion of severity variation. Moreover, comparison of severity percentages between periods shows that collisions occurring at night are associated with a 44\% higher risk of high-severity outcomes compared to those occurring during the afternoon. This pattern may be partially explained by traffic conditions. Collisions occurring during afternoon and evening periods are likely characterized by lower severity outcomes due to congested traffic conditions and reduced vehicle speeds, which limit impact forces and result predominantly in property-damage-only or minor injury collisions. On the flip side, night time collisions tend to exhibit higher severity rates despite lower traffic volumes, as uncongested conditions permit higher travel speeds, increasing the likelihood of severe injuries or fatalities. Additionally, reduced visibility, driver fatigue, and a greater prevalence of risky driving behaviors during late hours may further contribute to the elevated severity observed at night.

\begin{table}[ht]
\caption{Statistical Association Between Temporal Characteristics and Collision Severity}
\centering
\begin{tabularx}{\textwidth}{@{}>{\raggedright\arraybackslash}X
                            >{\centering\arraybackslash}X
                            >{\centering\arraybackslash}X
                            >{\centering\arraybackslash}X@{}}
\hline
\textbf{Temporal Factor} &
\multicolumn{3}{c}{\textbf{Statistical Results{$^{\mathrm{a}}$}}} \\
\cline{2-4}
 & \textbf{$\chi^2$ (df)} & \textbf{p-value} & \textbf{Cramér’s V} \\
\hline
Time of day & 146.29 (3) & $<0.001$ & 0.066 \\
Day of week & 13.34 (6)  & 0.038    & 0.020 \\
Month       & 45.89 (7)  & $<0.001$ & 0.037 \\
\hline
\multicolumn{4}{l}{$^{\mathrm{a}}$Statistical tests were applied to collision severity (High vs. Low) only.}
\end{tabularx}
\label{tab:severity_all}
\end{table}

Figure~\ref{fig3} shows the average number of collisions across days of the week alongside the percentage of high-severity collisions. Collision frequency remains relatively stable throughout the week, with slightly higher mean collision counts observed from Wednesday through Friday. Weekend days show a modest reduction in mean collision counts, with Sunday recording the lowest average number of collisions. Despite this reduction, the percentage of high-severity collisions is slightly higher on weekends, particularly on Sunday. The analysis of collision severity reveals a statistically significant association with day of the week (as shown in Table~\ref{tab:severity_all}). The chi-square test results indicate that high-severity percentages differ across weekdays, although the associated effect size is negligible (Cramér’s V = 0.020). This finding suggests that while small day-to-day differences in severity exist, the day of the week has minimal explanatory power with respect to collision severity. The elevated severity observed on weekends may be partly attributable to reduced traffic congestion, which allows for higher operating speeds and increases the probability of severe outcomes when collisions occur. In addition, weekend travel in Dubai often involves increased recreational trips, late-night activity, and longer-distance journeys on high-speed roadways, which may further contribute to elevated collision severity.

Figure~\ref{fig4} presents the mean daily number of traffic collisions by month together with the percentage of high-severity collisions. Collision frequency depicts moderate seasonal variation, with higher mean daily counts observed from February through May 2025, and a noticeable reduction in June 2025. However, the mean daily collision value for June is based on only two observations; therefore, it may not accurately represent the monthly average if all days were included. With respect to collision severity, the statistical results reveal a significant association between collision severity and month. Nevertheless, the corresponding effect size is small (Cramér’s V = 0.037), which indicates that monthly variation accounts for only a limited proportion of the observed differences in severity outcomes. The marginal effect of monthly variation observed in this study contrasts with findings from many other regions, where seasonal and weather related variability has been shown to exert a substantial influence on collision severity. In temperate and cold climate contexts, adverse conditions such as snowfall, fog, and reduced roadway friction have been associated with higher fatality risks and more severe injury outcomes~\citep{gorzelanczyk2025impact}. In contrast, Dubai experiences relatively stable climatic conditions throughout the year, with limited seasonal variation in precipitation, temperature, and visibility. This climatic stability likely reduces the influence of month-to-month variation on collision severity, which is consistent with the small effect size observed in this analysis.

\subsubsection{Pedestrian-Related Collision Incidents}
\begin{figure}[!ht]
\centering
\includegraphics[width=0.95\textwidth]{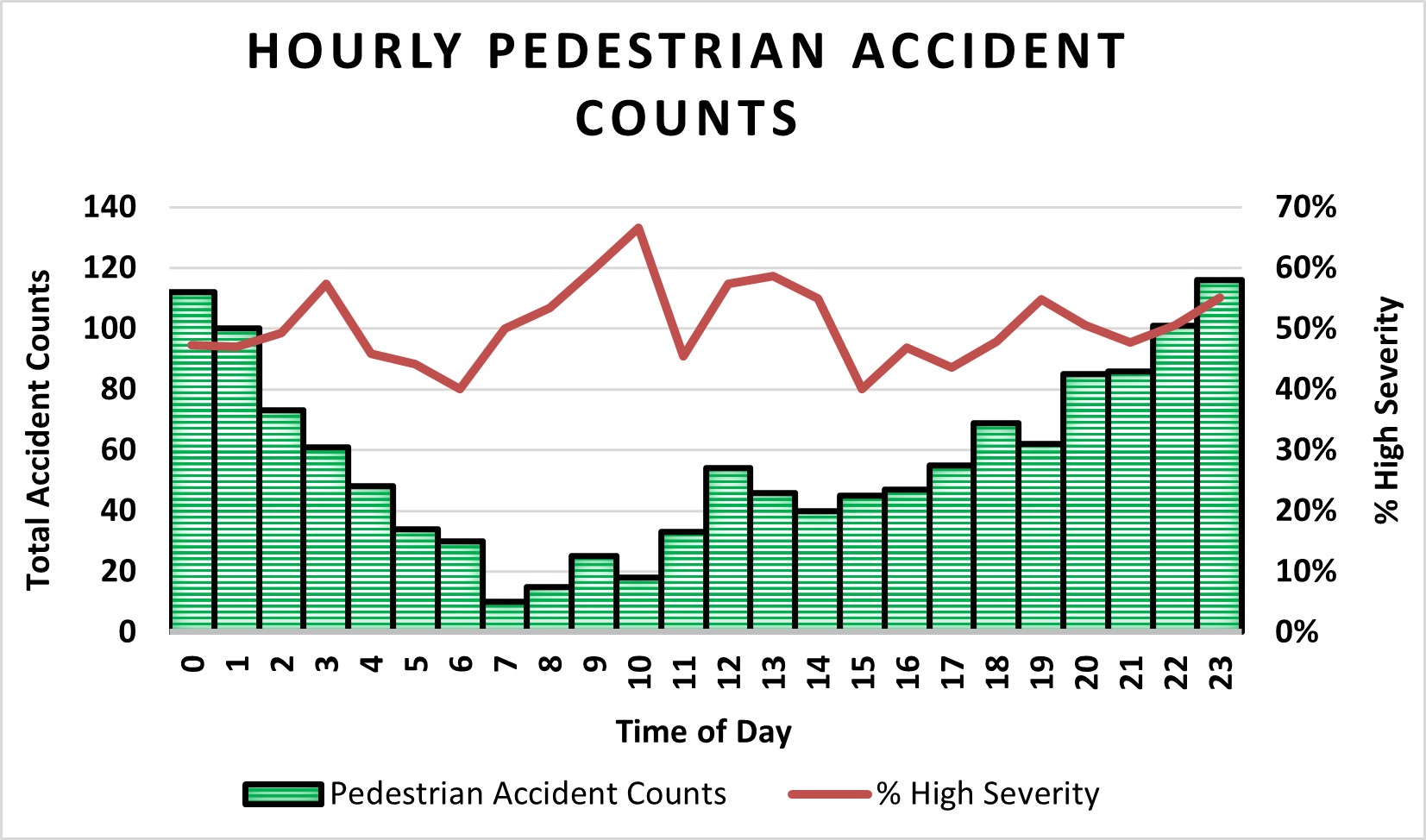}
\caption{Hourly distribution of pedestrian accidents.}
\label{fig5}
\end{figure}

\begin{figure}[!ht]
\centering
\includegraphics[width=0.95\textwidth]{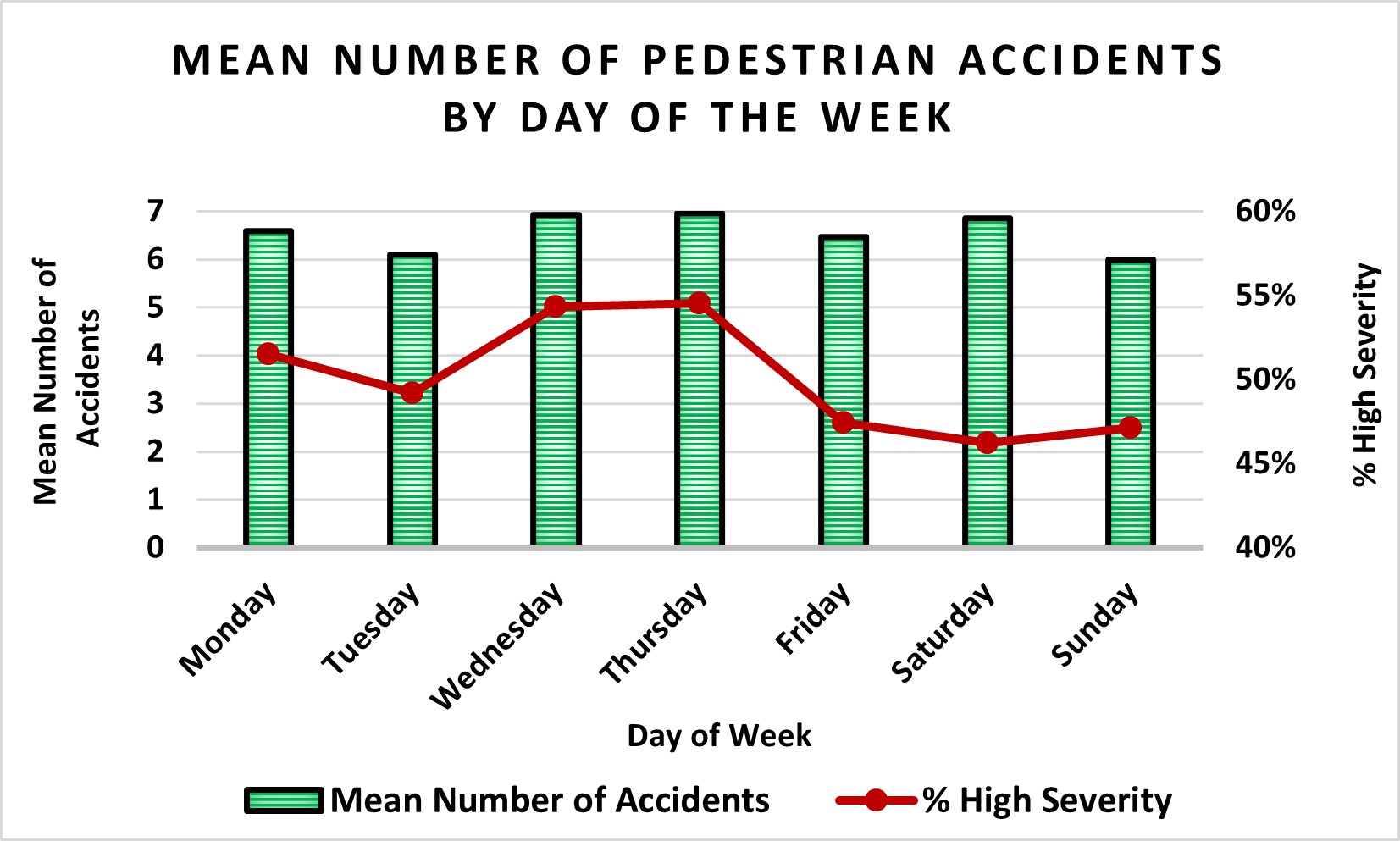}
\caption{Weekly distribution of pedestrian accidents.}
\label{fig6}
\end{figure}

\begin{figure}[!ht]
\centering
\includegraphics[width=0.95\textwidth]{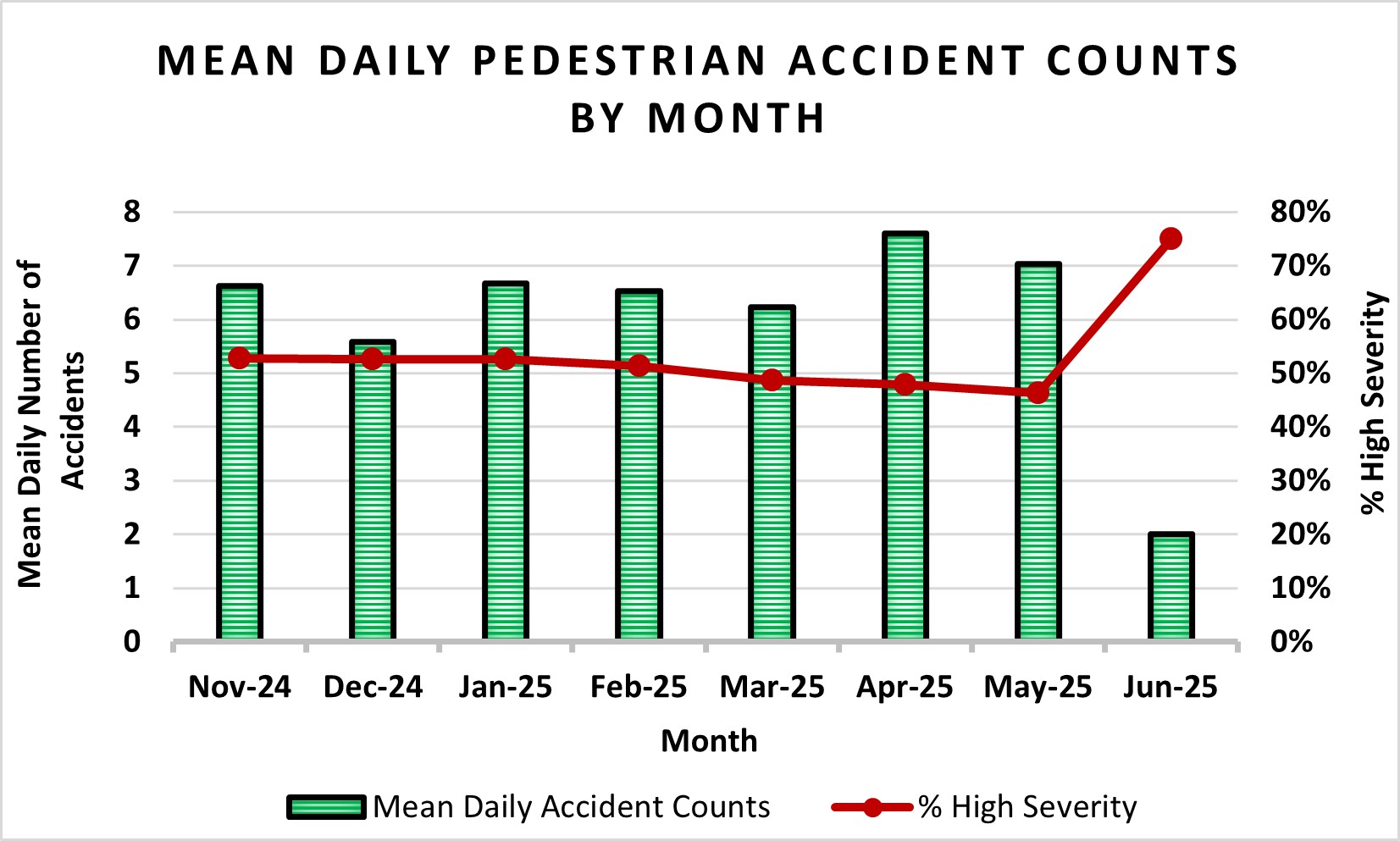}
\caption{Monthly distribution of pedestrian accidents.}
\label{fig7}
\end{figure}

The processed pedestrian accident dataset contains 1,365 records over the study period, yielding a mean of 6.56 pedestrian-related collisions per day in Dubai. The maximum daily pedestrian accident count occurred on March 28, 2025, with 16 incidents. In terms of severity, the highest number of high-severity pedestrian accidents on a single day was recorded on January 16, 2025, with 10 severe accidents. Temporal patterns at hourly, weekly, and monthly scales are presented in Figures~\ref{fig5}–\ref{fig7}.
\FloatBarrier

Figure~\ref{fig5} explains the hourly distribution of pedestrian accidents, showing a distinct temporal pattern. The highest pedestrian accident counts occur during the late evening hours, particularly between 23:00 and 00:00, and then decline steadily overnight, reaching their lowest levels during the early morning period (approximately 07:00–11:00). Thereafter, accident counts increase gradually throughout the day and reach the late-evening peak. 

Pedestrian accident severity exhibits a different temporal profile. The lowest percentage of sever pedestrian accidents occurs in the early morning between 6:00 and 7:00 (40\%) and in the afternoon between 15:00 and 16:00, while the highest percentage is observed in the late morning (10:00 - 11:00), reaching 66.7\%. Moreover, the percentage of sever accidents increases gradually from the late afternoon into late evening and generally decreases over the night. 

These differing trends may reflect variations in pedestrian exposure, traffic conditions, and visibility across the day. Higher accident frequencies during late evening hours are likely associated with increased pedestrian activity combined with reduced visibility for both drivers and pedestrians. The findings from virtual-simulation research by \cite{pala2024impact} reveal that low-light conditions may influence pedestrian decision-making, leading to riskier crossing behavior that could further explain the higher late-evening accident frequencies observed in this study. The elevated severity observed during late morning hours may be influenced by higher vehicle speeds under less congested traffic conditions, while lower severity during afternoon periods may reflect congested conditions in dense urban areas, where reduced operating speeds mitigate high severity outcomes. Despite these temporal variations, the statistical analysis presented in Table~\ref{tab:severity_pedestrian} indicates no significant association between pedestrian accident severity and aggregated time-of-day categories, with a negligible effect size (Cramér’s V = 0.023), which suggests limited explanatory power at this level of temporal aggregation.

\begin{table}[ht]
\caption{Statistical Association Between Temporal Factors and Pedestrian Accident Severity}
\centering
\begin{tabularx}{\textwidth}{@{}>{\raggedright\arraybackslash}X
                            >{\centering\arraybackslash}X
                            >{\centering\arraybackslash}X
                            >{\centering\arraybackslash}X@{}}
\hline
\textbf{Temporal Factor} &
\multicolumn{3}{c}{\textbf{Statistical Results{$^{\mathrm{a}}$}}} \\
\cline{2-4}
 & \textbf{$\chi^2$ (df)} & \textbf{p-value} & \textbf{Cramér’s V} \\
\hline
Time of day & 0.72 (3) & 0.869 & 0.023 \\
Day of week & 5.72 (6) & 0.455 & 0.065 \\
Month       & 2.96 (7) & 0.888 & 0.050 \\
\hline
\multicolumn{4}{l}{$^{\mathrm{a}}$Statistical tests were applied to pedestrian accident severity (High vs. Low) only.}
\end{tabularx}
\label{tab:severity_pedestrian}
\end{table}

As shown in Figure~\ref{fig6}, the average daily pedestrian accident counts remain mostly consistent across the week, with slightly higher mean values observed on Wednesday and Thursday and lower values toward the weekend, reaching a minimum on Sunday. The weekly distribution of high-severity pedestrian accidents follows a comparable pattern, and statistical testing (Table~\ref{tab:severity_pedestrian}) indicates that differences in the proportion of high-severity pedestrian accidents across days of the week are not statistically significant.  Similarly, Figure~\ref{fig7} shows that mean daily pedestrian accident counts remain relatively stable across months from November 2024 through May 2025, with a modest peak in April 2025 and a noticeable decline in June 2025. The proportion of high-severity pedestrian accidents also exhibits limited month-to-month variation, and the higher severity proportion observed in June 2025 is based on a very small number of observations and may not be representative. Consistent with the weekly results, statistical analysis confirms that monthly variations in pedestrian accident severity are not statistically significant.

These findings indicate that pedestrian accident occurrence and severity in Dubai are influenced by location and short-term temporal factors (i.e., time-of-day conditions) than by broader weekly or monthly cycles. This interpretation is supported by the spatial distribution of pedestrian accidents discussed in Section~\ref{sec:Results_Spatial}, which shows high concentrations in major urban centers along the northern and western coastal corridor. These areas are characterized by intense urban activity and consistently high pedestrian exposure throughout the week and across seasons, thereby limiting variability by day of the week or month. This may explain differences from the study by ~\cite{song2021mixed}, which reported elevated pedestrian injury severity during weekends in contexts where pedestrian risk is more influenced by weekend-specific behaviors such as late-night recreational travel and impaired driving.

\subsection{Spatial Hotspot Analysis}
This subsection examines the spatial hotspot analysis for all collision incidents and pedestrian-related collision accidents.

\subsubsection{All Collision Incidents}
\begin{figure}[!ht]
\centering
\includegraphics[width=0.95\columnwidth]{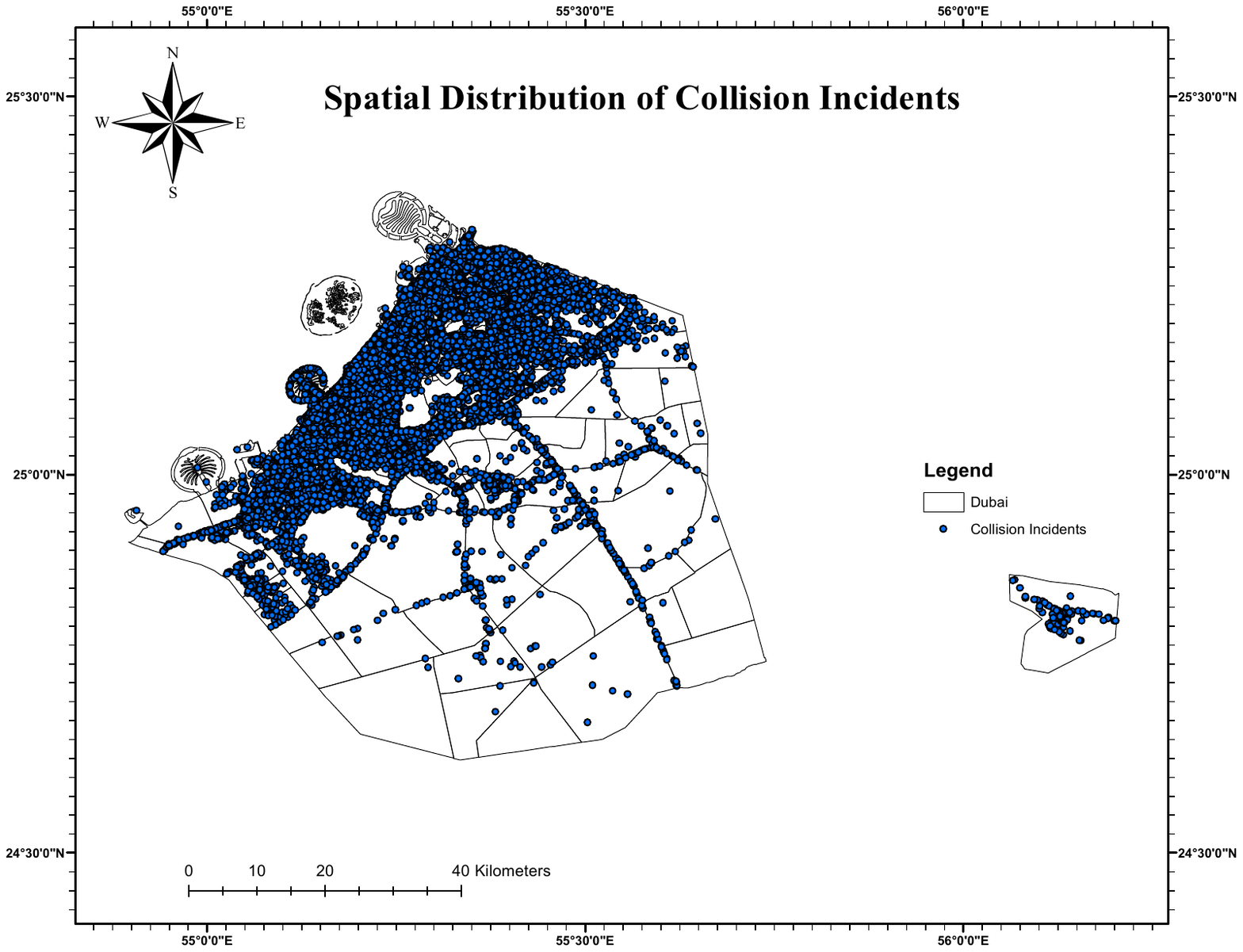}
\caption{Spatial distribution of collision incidents.}
\label{fig8}
\end{figure}

The spatial distribution of collision incidents across Dubai is presented in Figure~\ref{fig8}. The distribution of collisions shows a clear spatial heterogeneity, with collisions concentrated mainly in the northern, western, and northwestern parts of the city along the Arabian Gulf coastal corridor. These areas include major urban centers, dense residential and commercial developments, and primary transportation corridors, resulting in higher traffic exposure and interaction density. In contrast, the eastern and southern regions show comparatively lower collision points, corresponding to areas with lower traffic exposure and urban activity.

\begin{figure}[!ht]
\centering
\includegraphics[width=0.95\columnwidth]{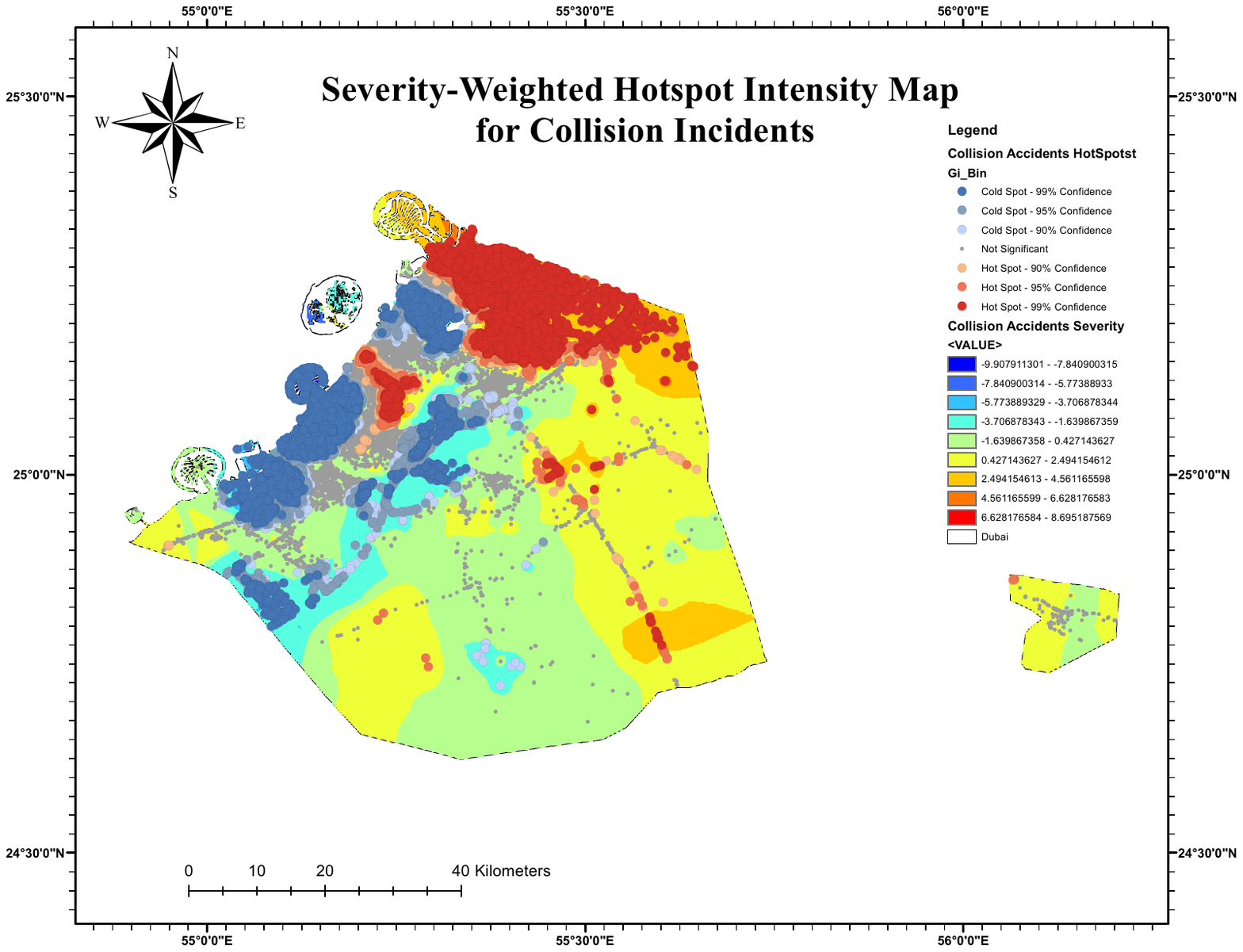}
\caption{Severity-weighted spatial hotspot analysis of collision incidents.}
\label{fig9}
\end{figure}

To identify statistically significant clusters of high-risk locations, a severity-weighted spatial hotspot analysis was conducted, and the results are illustrated in Figure~\ref{fig9}. The resulting hotspot intensity map identifies statistically significant hotspots (red color ramp) and cold spots (blue color ramp) at the 90\%, 95\%, and 99\% confidence levels. Statistically significant hotspots are primarily concentrated in the northern and northwestern parts of Dubai, as well as along the major highway connecting Dubai with Al Ain (Al Ain–Dubai Highway), indicating locations characterized by both frequent and severe collision incidents.

The northern and northwestern regions include major urban centers, dense residential and commercial developments, and primary transportation corridors. The presence of significant hotspots in these areas may be attributed to the concentration of high traffic volumes, complex roadway configurations, and major arterial corridors, combined with the coexistence of various vehicle types (e.g., buses, trucks, motorcycles, and passenger vehicles) which collectively increase collision energy and the probability of severe outcomes when accident occur. In addition, the Al Ain–Dubai Highway is a major intercity corridor characterized by high traffic volumes, frequent lane-changing maneuvers, and multiple merge and diverge points, which likely contribute to the observed hotspot patterns along this roadway.

Although the western parts of Dubai encompass a high concentration of collision incidents, statistically significant cold spots are also observed in portions of this region, which indicates areas characterized by relatively lower collision severity. These locations likely reflect roadway environments where traffic calming measures, congestion, or lower operating speeds reduce crash severity despite frequent incidents. In contrast, the eastern and southern regions are associated with statistically insignificant clusters, suggesting a random spatial distribution of collisions. 

\subsubsection{Pedestrian-Related Collision Incidents}
\label{sec:Results_Spatial}
\begin{figure}[!ht]
\centering
\includegraphics[width=0.95\columnwidth]{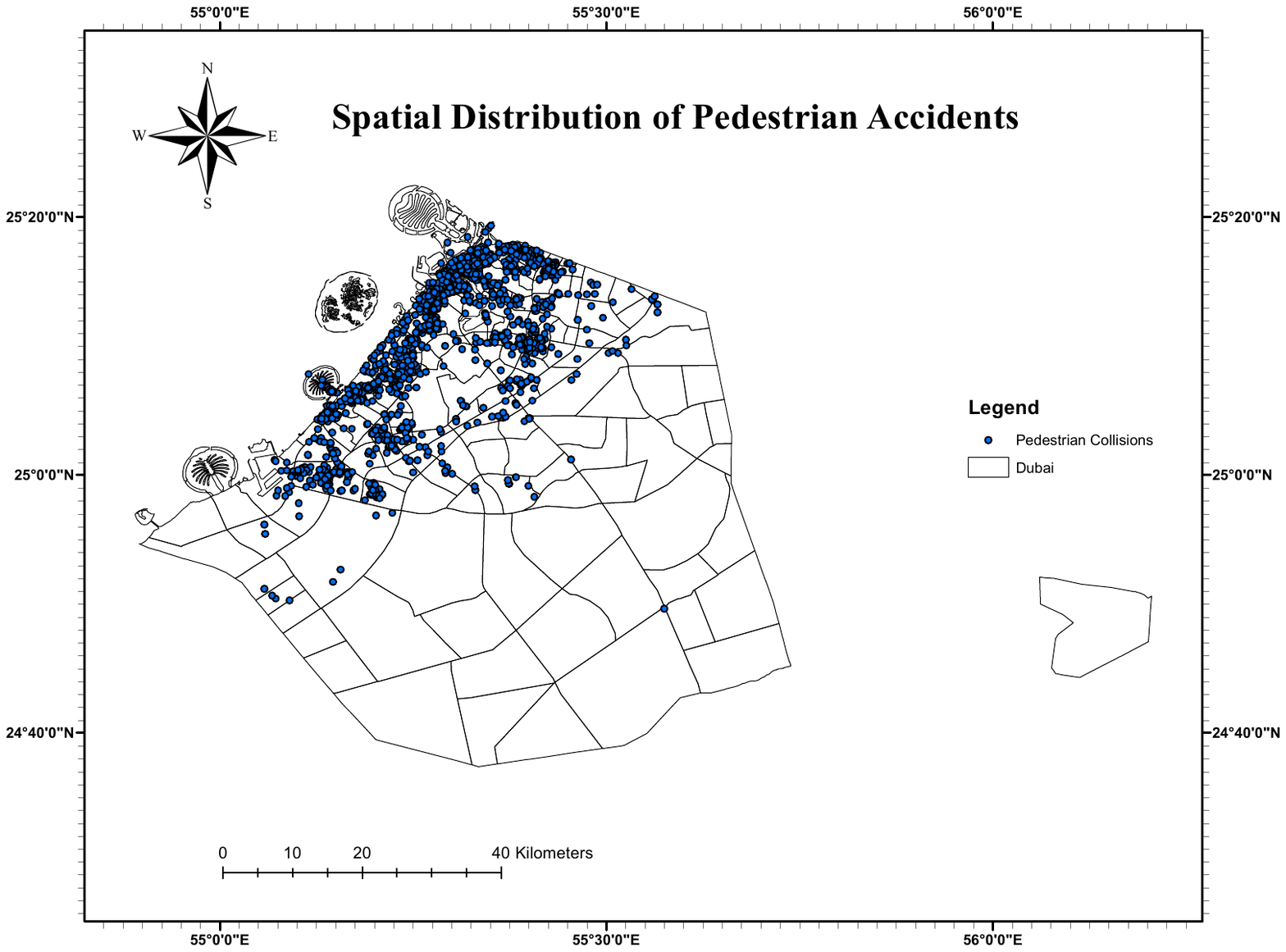}
\caption{Spatial distribution of pedestrian accidents.}
\label{fig10}
\end{figure}

Figure~\ref{fig10} depicts the spatial distribution of pedestrian accidents across Dubai. The distribution of pedestrian accidents shows a spatial pattern similar to that observed for collision incidents, with high concentrations in the northern, western, and northwestern parts of the city along the Arabian Gulf coastal corridor. This pattern is expected, as these areas encompass major urban centers, reflecting areas of intense urban activity and high pedestrian exposure.

\begin{figure}[H]
\centering
\includegraphics[width=0.95\columnwidth]{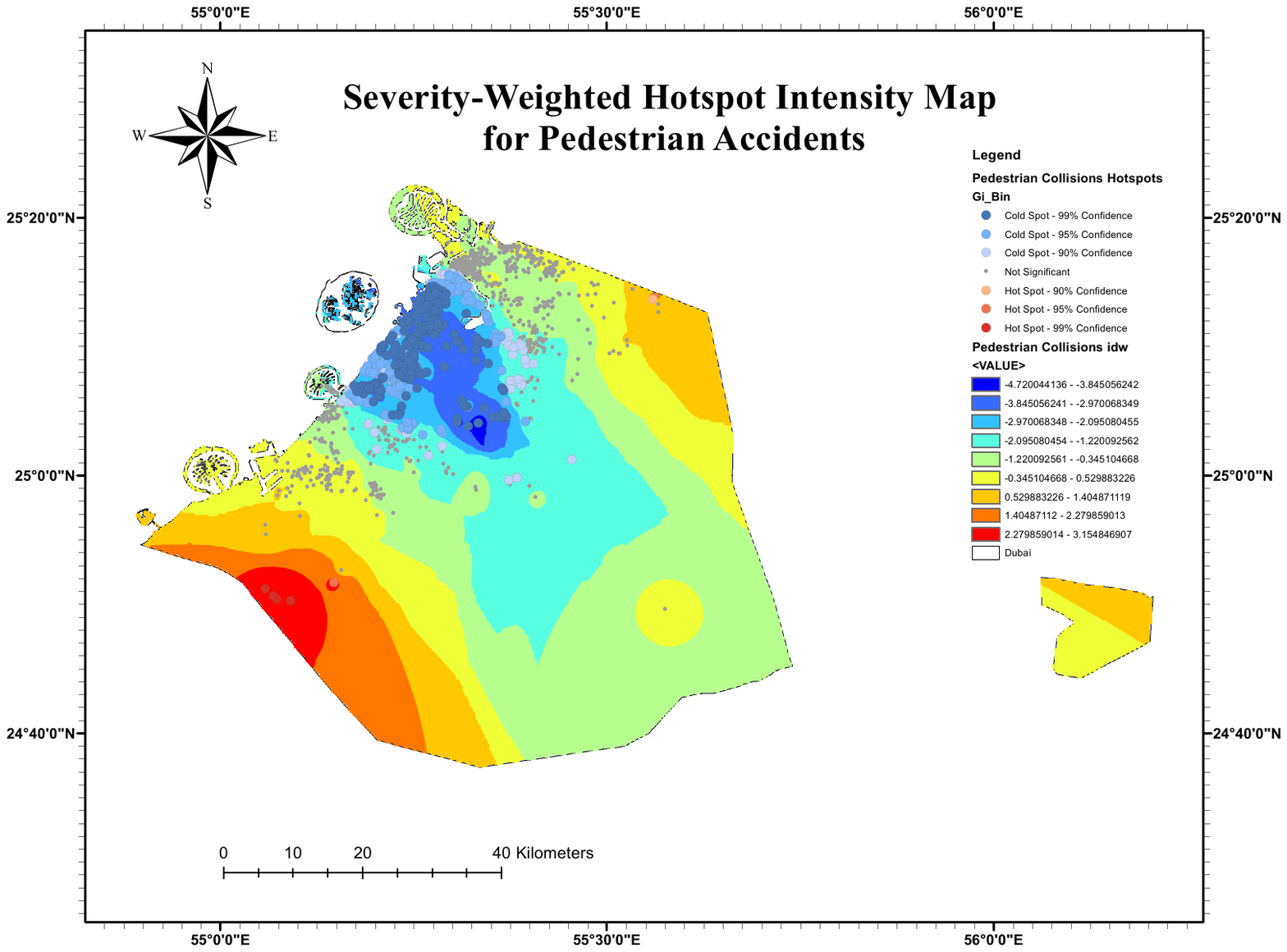}
\caption{Severity-weighted spatial hotspot analysis of pedestrian accidents.}
\label{fig11}
\end{figure}

The severity-weighted spatial hotspot analysis for pedestrian accidents is presented in Figure~\ref{fig11}. Statistically significant cold spots are identified in the central and northwestern regions, despite the relatively high concentration of pedestrian accidents in these areas. In major urban centers, pedestrian activity is typically associated with traffic calming measures, signalized intersections, lower speed limits, and recurrent congestion. Although pedestrian crashes may occur more frequently under these conditions, reduced operating speeds tend to reduce injury severity, which result in statistically significant cold spots.

Statistically significant hotspots are observed in the southwestern region, indicating clusters of high-severity pedestrian accidents despite comparatively lower accident frequencies. This pattern may be associated with the presence of Dubai Industrial City, which attracts employees and visitors and is surrounded by major roadways with unsignalized intersections (e.g., roundabouts) that allow continuous traffic movement and higher operating speeds. These conditions may increase the risk of severe pedestrian injuries when accidents occur, which point to the need for targeted traffic calming and pedestrian safety measures in this area. This finding is consistent with results reported by \cite{song2021mixed}, which indicate that industrial land-use areas are associated with an 8\% higher probability of pedestrian fatal injury. The remaining regions show statistically insignificant clusters, which suggests a random spatial distribution of pedestrian accidents across those areas.

\section{Discussion and Recommendations}
\label{sec: recomendations}
This section discusses the study findings and outlines the key recommendations that can be derived from the results.

\subsection{Discussion}
The temporal patterns observed in this study reflect the interaction between traffic and pedestrian demand, operating conditions, and user behavior. For overall collisions, higher frequencies during evening hours correspond to peak traffic demand, while the elevated severity observed during late-night and early-morning periods, despite lower collision counts, suggests that reduced congestion may enable higher operating speeds and, consequently, increase collision severity. Additional contributing factors may include reduced visibility, driver fatigue, and riskier driving behaviors during late hours. The spatial hotspot analysis reinforces these findings, identifying statistically significant severity hotspots in the northern and northwestern parts of Dubai and along major intercity corridors. These areas are characterized by dense urban activity, complex roadway configurations, and high-capacity arterial facilities, where heterogeneous traffic streams and frequent merging and turning movements increase the probability of high-energy collisions. Together, the temporal and spatial results indicate that severe collisions are concentrated in environments combining high traffic exposure with high operating speeds.

Pedestrian-related collisions show a different pattern. Although pedestrian accident frequency peaks during late evening hours, pedestrian injury severity does not vary significantly across broader temporal scales. Instead, pedestrian severity is more shaped by spatial context and roadway environment characteristics. The spatial analysis shows that pedestrian severity hotspots do not align with areas of highest pedestrian accident frequency. In dense urban centers, traffic calming, signalized crossings, and congestion probably reduce injury severity despite frequent pedestrian conflicts, whereas higher-speed environments with limited pedestrian infrastructure can produce severe outcomes even at lower crash frequencies.

Overall, the results demonstrate that collision severity arises from distinct mechanisms for overall collision incidents and pedestrian-involved crashes. Whereas overall collision severity reflects a combination of time-of-day operating conditions and high-risk roadway environments, pedestrian injury severity is shared primarily by spatial factors. These findings highlight the importance of context-sensitive analyses to support targeted and location-specific traffic safety interventions.

\subsection{Recommendations and Implications}
The results highlight the importance of adopting targeted and data-driven traffic safety policies that reflect temporal and spatial variations in collision severity across Dubai. The higher percentage of severe collisions during nighttime periods highlights the need for time-specific safety management strategies. Transportation agencies should consider implementing lower nighttime speed limits on selected corridors to account for reduced driver alertness, increased fatigue, and the higher operating speeds facilitated by uncongested traffic conditions during late hours. These measures should be complemented by enhanced automated enforcement, such as speed cameras and radar-based monitoring, to deter riskier driving behaviors that are more prevalent at night. Improvements in roadway lighting may further mitigate the elevated severity risks observed during nighttime periods.

The identification of statistically significant spatial hotspots of severe collisions in the northern and northwestern parts of the city, as well as along the Dubai–Al Ain Highway, highlights the importance of location-based intervention prioritization. These corridors would benefit from comprehensive road safety audits, targeted geometric design improvements at high-risk intersections, and upgraded traffic control devices. Focusing interventions on statistically significant hotspots, rather than applying uniform network-wide measures, can enhance the efficiency and effectiveness of safety investments.

Distinct spatial patterns in pedestrian accident severity further emphasize the need for severity-focused pedestrian safety interventions. High-severity pedestrian hotspots were observed in areas characterized by lower crash frequencies but more severe outcomes, often associated with high-speed arterial roadways and industrial land uses. In these environments, interventions such as pedestrian refuge islands, grade-separated pedestrian crossings on arterial corridors, localized speed limit reductions, physical traffic-calming measures, and enhanced roadway lighting should be prioritized to mitigate injury severity rather than focusing solely on reducing crash frequency.

Given the observed differences between collision frequency and severity patterns, transportation agencies are encouraged to adopt severity-weighted safety management practices. This includes regular updates of severity-weighted spatial hotspot analyses and continuous monitoring of statistically significant high-risk locations and time-of-day periods. Adopting such practices can support proactive, data-driven safety interventions and facilitate systematic evaluation of their effectiveness over time.

\section{Conclusion}
\label{sec:conclusions}
This study analyzed the temporal and spatial patterns of overall traffic collisions and pedestrian-related collisions in Dubai, UAE, with particular emphasis on collision severity. Temporal analyses examined hourly, weekly, and monthly variations in collision frequency and severity. The associations of temporal characteristics (i.e., time of day periods, day of week, and month) with collision severity were evaluated using chi-square tests and Cramér’s V. Spatial patterns were investigated through severity-weighted spatial hotspot analysis using the Getis-Ord Gi* statistic and inverse distance weighting (IDW) interpolation.

The results showed clear temporal variation in both the frequency and severity of overall traffic collisions. Higher collision frequencies were observed during evening and nighttime periods, from Wednesday through Friday, and between February and May. Pedestrian-related accidents showed a distinct temporal profile, characterized by higher occurrence during late-evening hours and relatively limited variation across days of the week and months, with slightly higher mean values observed on Wednesday and Thursday and in April 2025. Statistical analysis indicated that the association between temporal factors and collision severity was significant for overall collisions, though with small effect sizes. Notably, collisions occurring during nighttime periods were found to have a 44\% higher probability of being high-severity compared to those occurring during the afternoon. In contrast, no statistically significant associations were observed for pedestrian accident severity.

Spatial analysis demonstrated clear spatial heterogeneity in accident severity across Dubai. Severity hotspots for overall collisions were observed in the northern and northwestern parts of the city, and along the major highway connecting Dubai with Al Ain (Al Ain–Dubai Highway). Pedestrian severity hotspots, on the other hand, were primarily identified in the southwestern region, likely influenced by industrial land use and high-capacity roadways with unsignalized intersections. These findings provide valuable, data-driven insights to support targeted and location-specific traffic safety interventions.

The results support the adoption of targeted and data-driven traffic safety strategies. Time-based measures such as lower nighttime speed limits, enhanced automated enforcement, and improved roadway lighting are suggested for reducing severe nighttime collisions. Location-based interventions, including road safety audits, geometric design improvements, and pedestrian-focused treatments in identified statistically significant hotspots, can further enhance the effectiveness of safety investments.

This study has several limitations related to data availability. The traffic incident records did not include detailed information on contributing factors such as collision causation, vehicle speed at impact, driver or pedestrian age, vehicle type, or roadway functional class, which constrained the ability to evaluate underlying behavioral and roadway characteristics affecting collision frequency and severity. In addition, socio-economic and built-environment data at the city scale were not available for integration with the collision records, which limited the ability to evaluate the impact of these factors on collision outcomes. 

Future research should extend this framework to additional collision categories, such as vehicle–vehicle, rollover, bicycle, and animal-related accidents, to support a more comprehensive, data-driven approach to traffic safety planning in Dubai. Incorporating explanatory variables related to roadway design, traffic volumes, speed characteristics, land use, and environmental conditions would further strengthen causal interpretation.

\section*{Authorship contribution statement}
The authors confirm their contribution to the paper as follows: \textbf{Nael Alsaleh}.: Conceptualization, Methodology, Data curation, Investigation, Formal analysis, Software, Visualization, Supervision, Writing - original draft, and Writing - review \& editing. \textbf{Noura Falis}.: Data curation, Software, Visualization, Writing - original draft. \textbf{Tareq Alsaleh}.: Conceptualization, Methodology, Formal analysis, Visualization, Writing - original draft, and Writing - review \& editing. \textbf{Farah Ba Fakih}.:  Data curation, Writing - original draft.

\section*{Acknowledgements}
The authors would like to thank the Government of Dubai and the Dubai Police Department for making traffic incident data publicly available through the Dubai Pulse Open Data Platform.

\section*{Declaration of generative AI use}
AI tool (i.e., ChatGPT) was used only for grammar and language refinement, with no AI-generated original content. The authors reviewed and edited the content as needed and take full responsibility for the content of the published article.
\newpage
\bibliographystyle{abbrvnat}
\singlespacing
\bibliography{References.bib}
\end{document}